\documentclass[twocolumn,showpacs,superscriptaddress,preprintnumbers,amsmath,amssymb,prc]{revtex4-1}

\usepackage{xcolor}
\usepackage{graphicx}
\usepackage{dcolumn}
\usepackage{bm}
\usepackage{ifpdf}
\usepackage{epstopdf}
\usepackage{slashed}
\usepackage{amsfonts}
\usepackage{mathrsfs}
\usepackage{amssymb}
\usepackage{multirow}
\usepackage{CJK}
\usepackage{xcolor}
\usepackage{subfigure,dcolumn}
\usepackage[T2A,T1]{fontenc}
\usepackage[russian,english]{babel}
\usepackage{amsmath}
\usepackage{listings}
\usepackage{booktabs}
\begin{document}


\title{Machine learning the nuclear mass }

\author{Zepeng Gao}
\affiliation{College of Physics Science and Technology, Shenyang Normal University, Shenyang 110034, China}
\affiliation{School of Science, Huzhou University, Huzhou 313000, China}
\author{Yongjia Wang}
\email[Corresponding author, ]{wangyongjia@zjhu.edu.cn}
\affiliation{School of Science, Huzhou University, Huzhou 313000, China}
\author{Hongliang L\"{u}}
\affiliation{HiSilicon Research Department, Huawei Technologies Co., Ltd., Shenzhen 518000, China}


\author{Qingfeng Li}
\email[Corresponding author, ]{liqf@zjhu.edu.cn}
\affiliation{School of Science, Huzhou University, Huzhou 313000, China}
\affiliation{Institute of Modern Physics, Chinese Academy of Science, Lanzhou 730000, China}
\author{Caiwan Shen}
\affiliation{School of Science, Huzhou University, Huzhou 110000, China}

\author{Ling Liu}
\affiliation{College of Physics Science and Technology, Shenyang Normal University, Shenyang 110034, China}

\date{\today}

\begin{abstract}
\textbf{Background:} The masses of about 2500 nuclei have been measured experimentally, however more than 7000 isotopes are predicted to exist in the nuclear landscape from H (Z=1) to Og (Z=118) based on various theoretical calculations. Exploring the mass of the remains is a hot topic in nuclear physics. Machine learning has been served as a powerful tool in learning complex representations of big data in many fields.
\textbf{ Purpose:} We use Light Gradient Boosting Machine (LightGBM) which is a highly efficient machine learning algorithm to predict the masses of unknown nuclei and to explore the nuclear landscape in neutron-rich side from learning the measured nuclear masses.
\textbf{ Methods:} Several characteristic quantities (e.g., mass number, proton number) are fed into LightGBM algorithm to mimic the patterns of the residual $\delta(Z,A)$ between the experimental binding energy and the theoretical one given by the liquid-drop model (LDM), Duflo-Zucker (DZ) mass model, finite-range droplet model (FRDM), as well as the Weizs\"{a}cker-Skyrme (WS4) model, so as to refine these mass models.
\textbf{ Results:} By using the experimental data of 80 percent of known nuclei as the training dataset, the root mean square deviation (RMSD) between the predicted and the experimental binding energy of the remaining 20\% is about 0.234$\pm$0.022 MeV, 0.213$\pm$0.018 MeV, 0.170$\pm$0.011 MeV, and 0.222$\pm$0.016 MeV for the LightGBM-refined LDM, DZ, WS4, and FRDM models, respectively. These values are of about 90\%, 65\%, 40\%, and 60\% smaller than the corresponding origin mass models. The RMSD for 66 newly measured nuclei that appeared in AME2020 is also significantly improved on the same foot. One-neutron and two-neutron separation energies predicted by these refined models are in consistence with several theoretical predictions based on various physical models. In addition, the two-neutron separation energy of several newly measured nuclei (e.g., some isotopes of Ca, Ti, Pm, Sm) predicted with LightGBM-refined mass models are also in good agreement with the latest experimental data.
 \textbf{ Conclusions:} LightGBM can be used to refine theoretical nuclear mass models so as to predict the binding energy of unknown nuclei. Moreover, the correlation between the input characteristic quantities and the output can be interpreted by SHapley Additive exPlanations (SHAP, a popular explainable artificial intelligence tool), this may provide new insights on developing theoretical nuclear mass models.

\end{abstract}

\maketitle

\section{Introduction}


The mass of nuclei, which is of fundamental importance to explore nuclear landscape and properties of nuclear force, plays a crucial role in understanding many issues in both the fields of nuclear physics and astrophysics \cite{1,2,3,4,Niu:2018olp,44}. It is known that more than 7000 nuclei in the nuclear landscape from H (Z=1) to Og (Z=118) are predicted to be existed according to various theoretical models, while about  3000 nuclei have been found or synthesized in experimental and the masses of about 2500 nuclei have been measured accurately \cite{5,6}. Exploring the masses of the remains is of particular interest for both nuclear experimental and theoretical community. On the experimental side, facilities, such as HIRFL-CSR in China, RIBF at RIKEN in Japan, cooler-storage ring ESR and SHIPTRAP at GSI in Germany, CPT at Argonne and LEBIT at MSU in US, ISOLTRAP at CERN, JYFLTRAP at Jyv\"{a}skyl\"{a} in Finland, TITAN at TRIUMPH in Canada, are partly dedicated to measuring the nuclear mass, especially for nuclei round the driplines. On the theoretical side, various models have been developed to study nuclear mass by considering different physics, such as finite-range droplet model (FRDM) \cite{7,8}, the Weizs\"{a}cker-Skyrme (WS) model \cite{9}, Hartree-Fock-Bogoliubov (HFB) mass models \cite{10,11,12}, the relativistic mean-field (RMF) model \cite{Geng:2005yu}, relativistic continuum Hartree-Bogoliubov (RCHB) theory \cite{Xia:2017zka}. Though tremendous progress has been made in both experimental and theoretical sides, exploring mass of nuclei around dirplines is still a great challenge for both sides.

Machine learning which is the subset of artificial intelligence has been widely applied for analyzing data in many branches of science, such as in physics, e.g., Refs. \cite{ROMP,Nature1}. In nuclear physics, a Bayesian neural network (BNN) has been applied to reduce the mass residuals between theory and experiment, and a significant improvement in the mass predictions of several theoretical models was obtained after BNN refinement \cite{Utama:2015hva,Utama:2017wqe,Niu:2018csp}, e.g., the root mean square deviation (RMSD) of the liquid-drop model (LDM) was reduced from about 3 MeV to 0.8 MeV. Later on, BNN approach is also applied to study nuclear charge radii \cite{Utama:2016tcl}, $\beta$-decay half-lives \cite{Niu:2018trk}, fission product yield \cite{PJC}, fragment Production in spallation reaction \cite{Ma:2020bic,Ma:2020mbd}. Besides BNN, other machine learning or deep learning algorithms also have been employed in studying of nuclear reactions, e.g., Refs. \cite{PLG,Du:2019civ,Steinheimer:2019iso,Song:2021rmm,Wang:2020tgb,Li:2020qqn}. Focusing on nuclear mass, besides BNN in Refs. \cite{Utama:2015hva,Utama:2017wqe,Niu:2018csp}, the Levenberg-Marquardt neural network approach \cite{Zhang:2017zvb}, Gaussian processes \cite{shelley,Neufcourt:2018syo}, decision tree algorithm \cite{Carnini:2020lvr}, the Multilayer Perceptron (MLP) algorithm \cite{MLP} also have been applied to refine nuclear mass models.

Indeed, studying nuclear mass with machine learning algorithms is not a new topic and it can be traced back to at least 1993, see e.g., Refs. \cite{Gernoth:1993dqa,Athanassopoulos:2003qe,Clark:2006ua} and reference therein. In Ref. \cite{Gernoth:1993dqa}, the capability of multilayer feedforward neural networks for learning the systematics of atomic masses and nuclear spins and parities with high accuracy has been found. This topic is flourishing again because of the rapid development of computer science and artificial intelligence. In 2016, Light Gradient Boosting Machine (LightGBM) which is a tree based learning algorithm was developed by Microsoft \cite{LightGBM}. It is a state-of-the-art machine learning algorithm which has achieved better performances in many machine learning tasks. Therefore, it would be interesting to explore whether LightGBM algorithm can achieve better accuracy than BNN on the task of predicting nuclear mass.

The paper is organized as follows: In Sec.\ref{LightGBM and the input features}, we will introduce the LightGBM model and ten input features. The predicted binding energy and neutron separation energy obtained with LightGBM are discussed in detail in Sec.\ref{sec:artwork}. The conclusion and outlook are given in Sec.\ref{Conclusion}.

\section{LightGBM and the input features}
\label{LightGBM and the input features}

LightGBM refers to a recent improvement of gradient boosting
decision tree (GBDT) that provides
efficient implementation of gradient boosting algorithms.
It is becoming more popular by the day due to its efficiency
and capability of handling large amounts of data. LightGBM
has leaf-wise growth of trees, rather than a level-wise
growth. After the first partition, the next split is
performed only on the leaf node that adds more to the
information gain.

The primary advantage of LightGBM is the change in training
algorithm that speeds up the optimization process dramatically
and results in a more effective model in many cases. More
concretely, to speed up the training process, LightGBM uses
a histogram-based methodology to select the best segmentation.
For any continuous variable, instead of using individual values,
these are divided into bins or buckets, which can accelerate
the training process and reduce memory usage. In addition, LightGBM contains two novel techniques: Gradient
bases One-Side Sampling (GOSS), which keeps all the instances
of large gradient and performs random sampling on the instances
with small gradient, and Exclusive Feature Bundling (EFB),
which helps to bundle multiple features into a single feature
without losing any information. Furthermore, as a decision tree-based algorithm, LightGBM also has high level of interpretability, allowing the results obtained in machine learning model to be checked against previous knowledge regarding nuclear mass. For example, one can find which feature is more important for predicting nuclear mass, this would be helpful to further improve nuclear mass model.

In this work, the binding energies of 2408 nuclei between $^{16}$O and $^{270}$Ds from the atomic mass evaluation (AME2016) \cite{6} are employed as the training and testing dataset. LightGBM is trained to learn the residual between the theoretical prediction and the experimental binding energy, $\delta(Z,A){\rm{ }} = {\rm{ }}{B_{{\rm{th}}}}(Z,A) - {B_{{\rm{exp}}}}(Z,A)$. Four theoretical mass models are adopted in this work to obtain $B_{th}$, including the LDM \cite{Zhang:2017zvb}, DZ \cite{Duflo:1995ep}, FRDM \cite{7,8}, and WS4 \cite{9}. After LightGBM learns the behaviour of the residual $\delta(Z,A)$, the binding energy of an unknown mass nucleus can be obtained via $B_{\rm{LightGBM}}(Z,A)$=$B_{\rm{th}}(Z,A)$+$\delta(Z,A)$. It is found that the RMSD of these four theoretical mass models can be significantly improved after LightGBM refinement.

For the LDM model, nucleus
is regarded as a non-compressible droplet, which
contains the volume energy, surface energy, Coulomb
energy of proton repulsion, the symmetry energy related
to the ratio of neutrons to protons, and the pairing
energy of the neutron-proton pairing effect. It can be
described as follows:
\begin{equation}
\begin{aligned}
B_{\mathrm{LDM}}(Z, A)= \;&a_{v}\left(1+\frac{4 k_{v}}{A^{2}} |T_{z}|\left(|T_{z}|+1\right)\right) A \\
&+a_{s}\left(1+\frac{4 k_{s}}{A^{2}} |T_{z}|\left(|T_{z}|+1\right)\right) A^{\frac{2}{3}} \\
&+a_{c} \frac{Z^{2}}{A^{\frac{1}{3}}}+f_{p} \frac{Z^{2}}{A}+E_{p}.
\end{aligned}
\end{equation}
Where $E_p$ is the pairing energy given by the following
expression:
\begin{equation}
E_p=
\left\{
\begin{aligned}
&\frac{d_n}{N^{1/3}}+\frac{d_p}{Z^{1/3}}+\frac{d_{np}}{A^{2/3}},&{\rm{for}\ \textit{Z}\ \rm{and}\ \textit{N}\ \rm{odd,}} \\
&\frac{d_p}{Z^{1/3}},&{\rm{for}\ \textit{Z}\ \rm{odd,}\ \textit{N}\ \rm{even,}} \\
&\frac{d_n}{N^{1/3}},&{\rm{for}\ \textit{Z}\ \rm{even,}\ \textit{N}\ \rm{odd,}} \\
&0,&{\rm{for}\ \textit{Z}\ \rm{and}\ \textit{N}\ \rm{even.}} \\
\end{aligned}
\right.
\end{equation}

In the above formula, \textit{A}, \textit{Z}, \textit{N} and
$T_z$ are the mass number, proton number, neutron number and
the third component of isospin $(T_z =\frac{1}{2}(Z-N))$, $a_v$, $k_v$, $a_s$, $k_s$, $a_c$, $f_p$, $d_n$, $d_p$, $d_{np}$ are adjustable
parameters with the values given in Table \ref{tab:parameters}. Based on these parameters the binding energy of 2408 nuclei
theoretical and experimental values as an RMSD is $2.463$ MeV.

\begin{table}[!t]
\caption{Parameter setting}
\label{tab:parameters}
\begin{tabular*}{8cm} {@{\extracolsep{\fill} } llr}
\toprule

Parameter & Value (MeV) \\
\midrule
$a_v$  & -15.4963 \\
$a_s$   & 17.7937      \\
$k_v$   & -1.8232   \\
$k_s$  & -2.2593\\
$a_c$ & 0.7093\\
$f_p$ & -1.2739 \\
$d_n$ & 4.6919 \\
$d_p$ & 4.7230 \\
$d_{np}$ & -6.4920 \\
\bottomrule
\end{tabular*}
\end{table}

This work mainly aims to find the relationship between the
feature quantity of each nucleus and $\delta(Z,A)$ with the
LightGBM model. For each nucleus, we selected 10 physical quantities
(cf. Table \ref{tab:features}) as the input features. It is known that nuclear binding energy and nuclear structure are linked, therefore,
we selected four physical quantities related to the shell
structure, among which $Z_m$ and $N_m$ are the shells where
the last proton and neutron are located, and the level of
the shell is given by the magic numbers. The number
of protons between 8, 20, 50, 82 and 126 corresponds to
$Z_m$ of 1, 2, 3, 4, and the number of neutrons between
8, 20, 50, 82, 126 and 184 corresponds to $N_m$ of 1, 2, 3,
4, 5. In addition, $|Z - m|$ and $|N - m|$ are the absolute
values of the difference between the number of protons,
the number of neutrons, and the nearest magic number,
respectively, which represent the distance between the
number of protons, the number of neutrons, and the
nearest magic number. N pair is an index that considers
the proton-neutron pairing effect, the odd-odd nucleus is
0, the odd-even nucleus or even-odd nucleus is 1, and the
even-even nucleus is 2.

\begin{table}[!t]
\caption{Selection of characteristic quantities}
\label{tab:features}
\begin{tabular*}{8cm} {@{\extracolsep{\fill} } llr}
\toprule

Features & Description \\
\midrule
A  & mass number \\
Z   & proton number      \\
N   & neutron number   \\
N/Z   & ratio of neutron to proton \\
$B_{LDM}$ & theoretical value from LDM \\
$N_{pair}$ = 0,1,2  & dependence on pair effect; \\
& for odd-odd,odd-even,even-even\\
$Z_m$ = 1,2 $\cdots$ & shell of the last proton;\\
 & for $8\leqslant Z < 20$,$20\leqslant Z < 28$,$\cdots$\\
$N_m$ = 1,2 $\cdots$ & shell of the last neutron;\\
 &  for $8\leqslant N < 20$,$20\leqslant N < 28$,$\cdots$\\
$|Z-m|$ & the distance between the proton number \\
& and the nearest magic number; \\
& $m \in \left\{8,20,28,50,82,126\right\}$ \\
$|N-m|$ & the distance between the neutron number \\
& and the nearest magic number; \\
& $m \in \left\{8,20,28,50,82,126,184\right\}$ \\

\bottomrule
\end{tabular*}
\end{table}

In this work, the value of num\_boost\_round (maximum number of decision trees allowed) is 50000, num\_leaves (maximum number of leaves allowed per tree) is 10, max\_depth (maximum depth allowed per tree) is -1 and other parameters are basically set as their default values of the LightGBM model. Varying these parameters would not alter the results significantly. During the training process,
LightGBM will generate a decision tree based on the
relevant information between the features of the training
set and $\delta (Z, A)$. 10-fold cross-validation, which is a technique to evaluate models by partitioning the original dataset into 10 equal size subsamples., is also applied to prevent overfitting and selection bias. After training, the model will make
predictions on the testing set. Each nucleus in the testing
set traverses the decision tree grown during model
training. Each decision tree will give its contribution to
the predicted value according to the feature quantity of
each nucleus. The sum of the contributions of all decision
trees is the predicted value given by the final model.

\section{Result}\label{sec:artwork}

\subsection{Predictions on the binding energy based on LDM}

\begin{figure*}
    \centering
    \includegraphics[width=\textwidth]{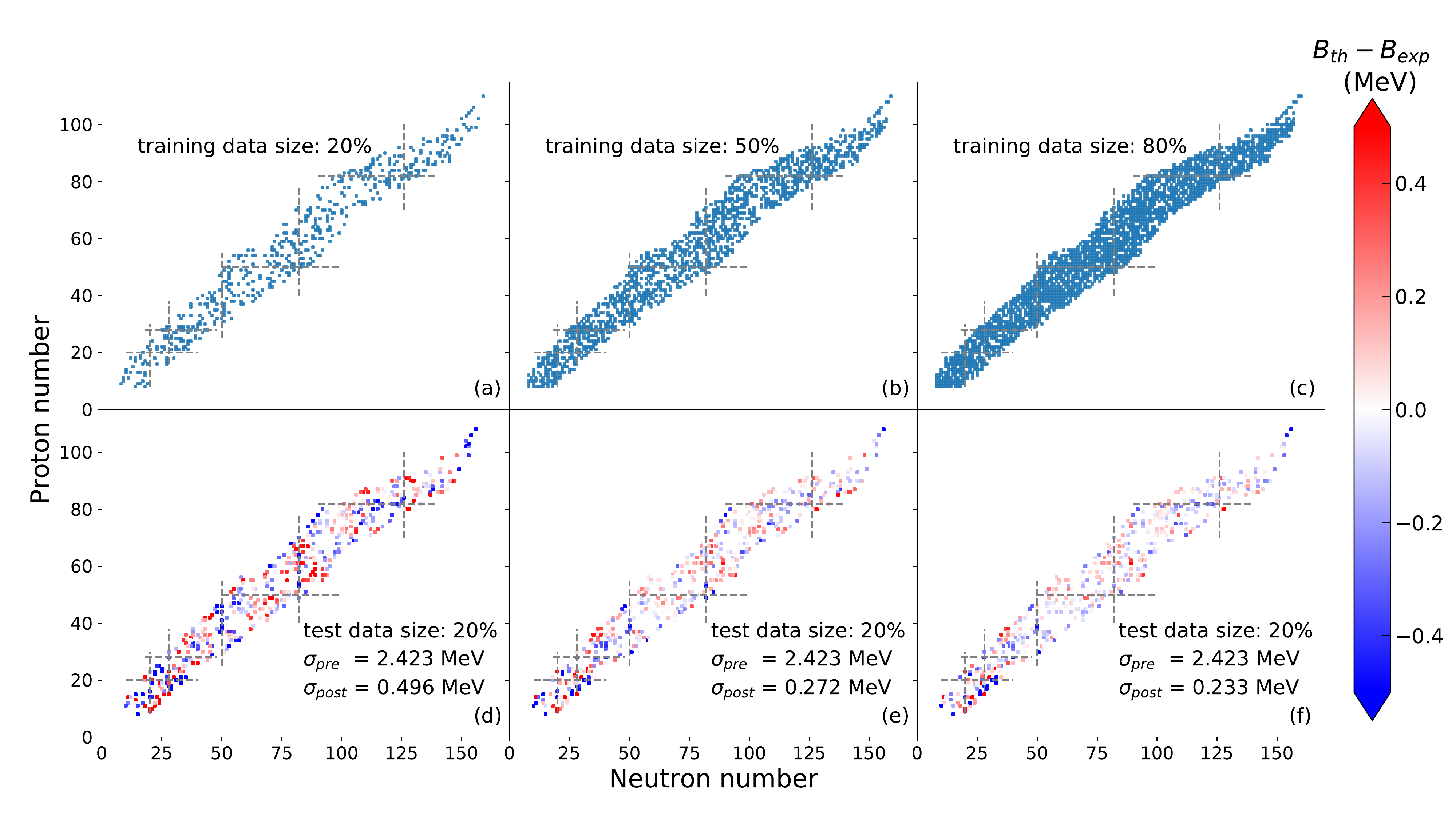}
    \caption{Upper panels: Locations of training data sets with 20\% (a), 50\% (b), and 80\% (c) of the 2408 nuclei from AME2016 in the N-Z plane. Lower panels: The absolute error between the experimental and LightGBM-refined LDM predicted binding energies for the testing set (20\% of the 2408 nuclei). (d), (e), and (f) display the deviation obtained with LightGBM-refined LDM trained with 20\%, 50\%, and 80\% of the 2408 nuclei. $\sigma_{pre}$ is RMSD of the original LDM, $\sigma_{post}$ is RMSD of the LightGBM-refined LDM. }
    \label{fig:2}
\end{figure*}

\begin{figure}[hbt!]
    \centering
    \includegraphics[width=\linewidth]{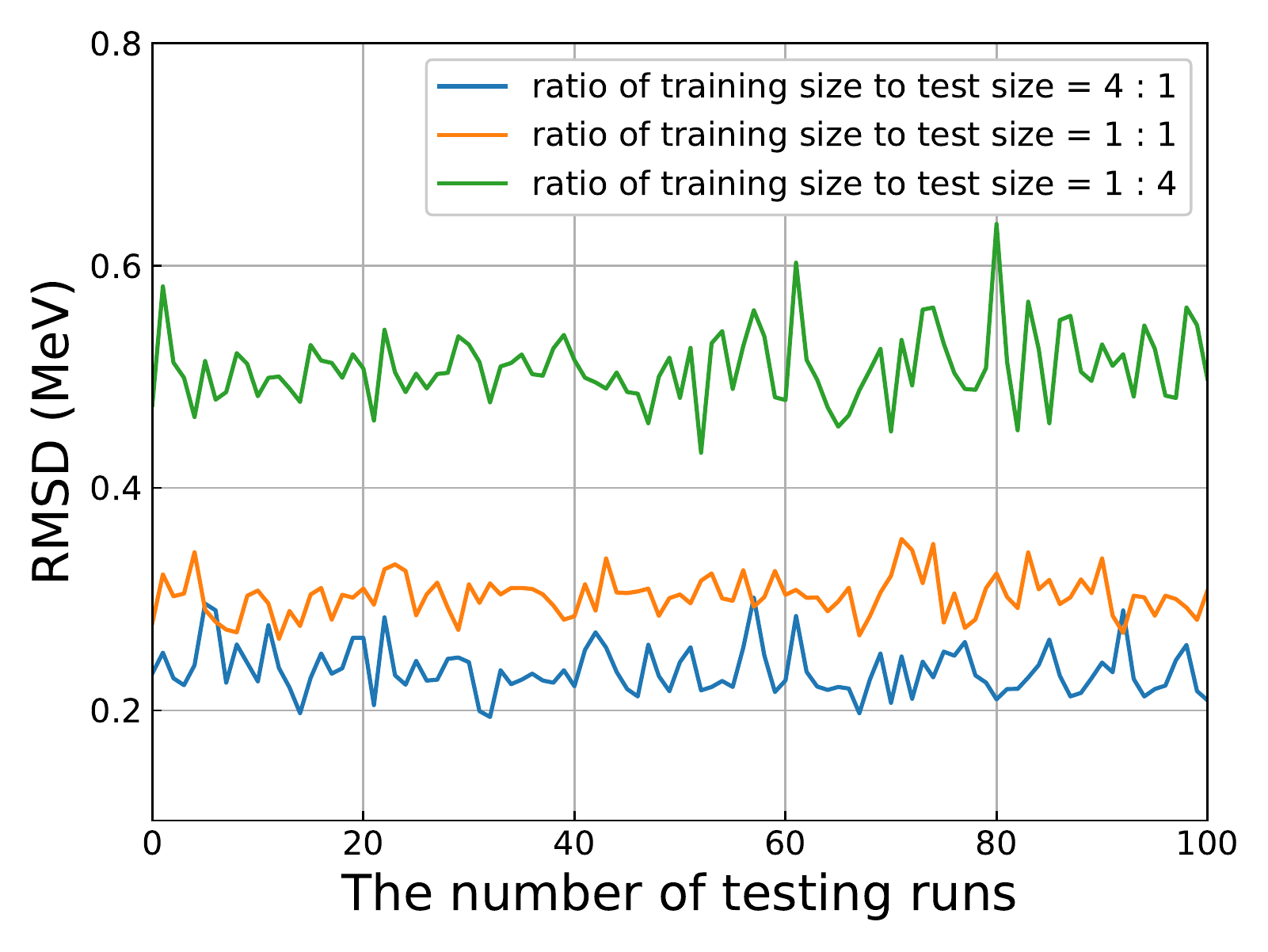}
    \caption{ The RMSD for the testing data from 100 runs. In each run, the 2408 nuclei are randomly split into the training and testing data sets with the ratio of 4:1 (blue), 1:1 (orange), and 1:4 (green).}
    \label{fig:3}
\end{figure}
\begin{figure}[hbt!]
    \centering
    \includegraphics[width=\linewidth]{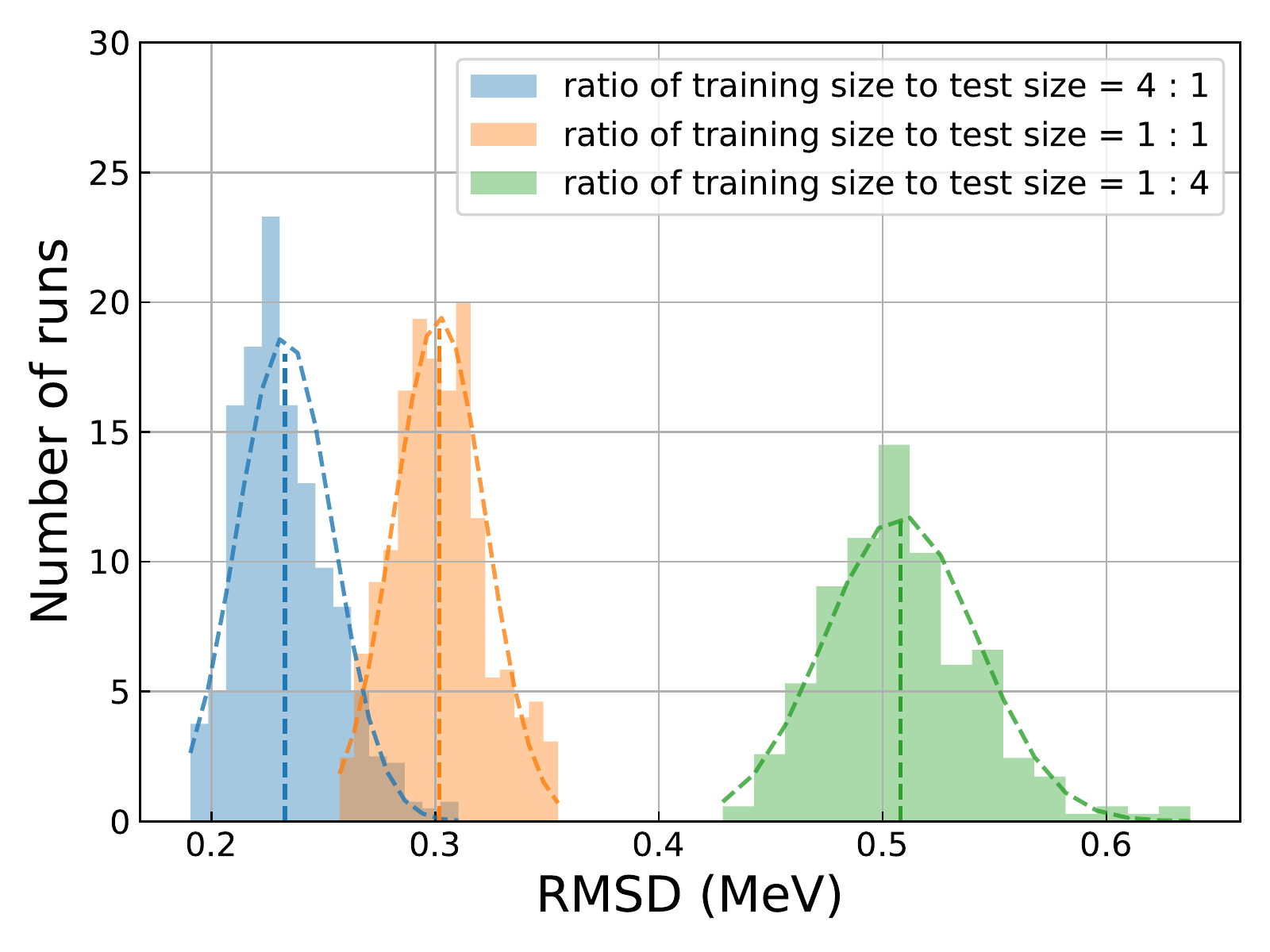}
    \caption{The density distribution of RMSD between the experimental and the predicted binding energy. Results from 500 runs for each set is displayed. Dash lines denote Gaussian fit to the distribution. The mean values and the standard deviation of RMSD are 0.508, 0.303, 0.224 MeV and 0.035, 0.020, 0.022 MeV for the three sets with different ratios of training to testing size, respectively. }
    \label{fig:4}
\end{figure}

\begin{figure*}
    \centering
    \includegraphics[width=0.8\textwidth]{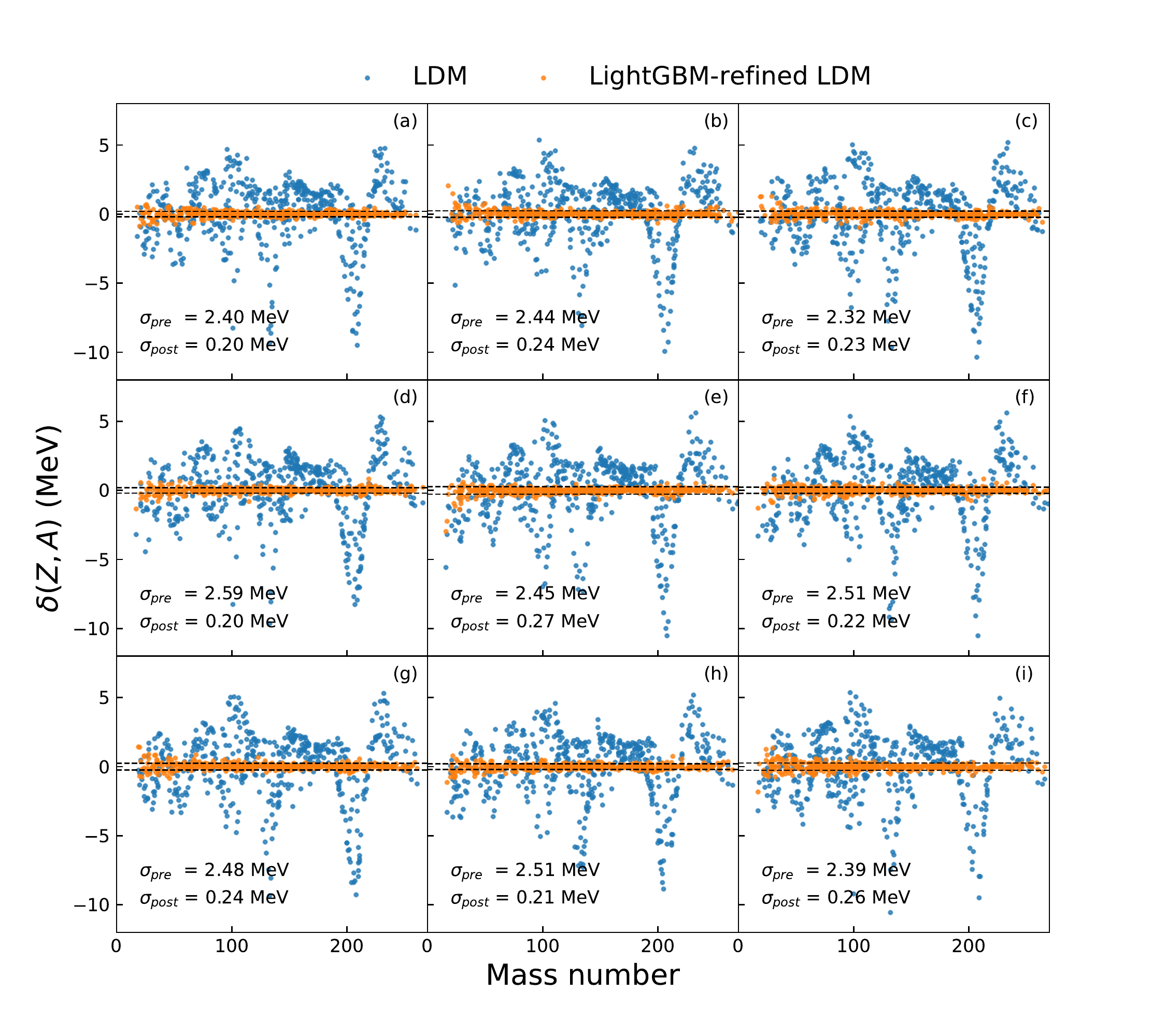}
    \caption{The residual $\delta(Z,A)$ is plotted as a function of mass number. Nine runs with randomly splitting the 2408 nuclei into the training and testing groups with the ratio of 4:1 are displayed. Blue
    and orange points denote $\delta(Z,A)$ for testing data obtained with the LDM and the LightGBM refined LDM, respectively. $\sigma_{pre}$ is RMSD of the original LDM, $\sigma_{post}$ is RMSD of the LightGBM-refined LDM. }
    \label{fig:5}
\end{figure*}
In this section, LightGBM is trained to learn the residual $\delta(Z,A)$ between the LDM and experimental binding energies. For this purpose, binding energies of the 2408 nuclei between $^{16}$O and $^{270}$Ds from AME2016 are split into training and testing data sets. We note here that, nuclei with proton (neutron) number smaller than 8 and with relatively large experimental uncertainties in AME2016 are not used. First,  the influence of training size on the predicted binding energy is examined, as displayed in Fig.\ref{fig:2}. We randomly select 482 (about 20\% of the 2408 nuclei) nuclei to constitute the training set. The RMSD of LDM for the selected 482 nuclei is about 2.458 MeV, after the LightGBM refinement, the RMSD is reduced to 0.496, 0.272, 0.233 MeV when 482, 1204, and 1926 nuclei are used to train the LightGBM, respectively. This means that LightGBM has been able to capture the missing physics of the LDM and to decode the correlation between the input features and the residual, so as to further improve the agreement with experimental data.

In addition, it can be seen that the deviation between the experimental and LightGBM-refined LDM predictions for nuclei with the small number of proton and neutron is usually larger, this may due to the fact that the microstructure effect in the light-mass nuclei is strong, and there are less data on the light-mass nuclei in the training set. The value of the above mentioned RMSD will fluctuate when the training and test data sets are randomly selected, because $\delta(Z,A)$ for some of nuclei (i.e., nuclei around magic number) are large and some of are small. To evaluate this issue, we randomly split the 2408 nuclei into training and testing data sets 500 times with each ratio (i.e., 4:1, 1:1, and 1:4), the RMSD and its density distribution are plotted in Fig.\ref{fig:3} and Fig.\ref{fig:4}. As observed in Fig.\ref{fig:3}, fluctuation in the RMSD is the largest of all when the ratio of training size to testing size is 1:4. The RMSD for 1926 nuclei predicted by LightGBM-refined LDM with learning the binding energy of 482 nuclei is about 0.508$\pm$0.035 MeV, this result is comparable to many physical mass models. With the training data set is built from 1926 nuclei and the remaining 482 nuclei constitute the testing data set, the RMSD is obtained to be 0.234$\pm$0.022 MeV, this performance is better than many physical mass models.

Fig.~\ref{fig:5} shows the residual $\delta(Z,A)$ obtained from the LDM and LightGBM-refined LDM. The results from nine runs with randomly selected 80\% of 2408 nuclei as the training set and the remaining 20\% as the testing set are displayed. It can be seen that the residual $\delta(Z,A)$ obtained with the original LDM is large, especially for nuclei around magic number, due to the absence of shell effect in the LDM. After the refinement of LightGBM, $\delta(Z,A)$ is considerably reduced, especially for nuclei with mass number larger than 60. The performance of LightGBM for nuclei with mass number smaller than 60 is not as good as that for nuclei with large mass number, the same as we already observed in Fig.\ref{fig:2}. This could be improved by feeding more relevant features to LightGBM.

\subsection{Predictions on the binding energy based on different mass models}

\begin{figure*}
    \centering
    \includegraphics[width=0.8\textwidth]{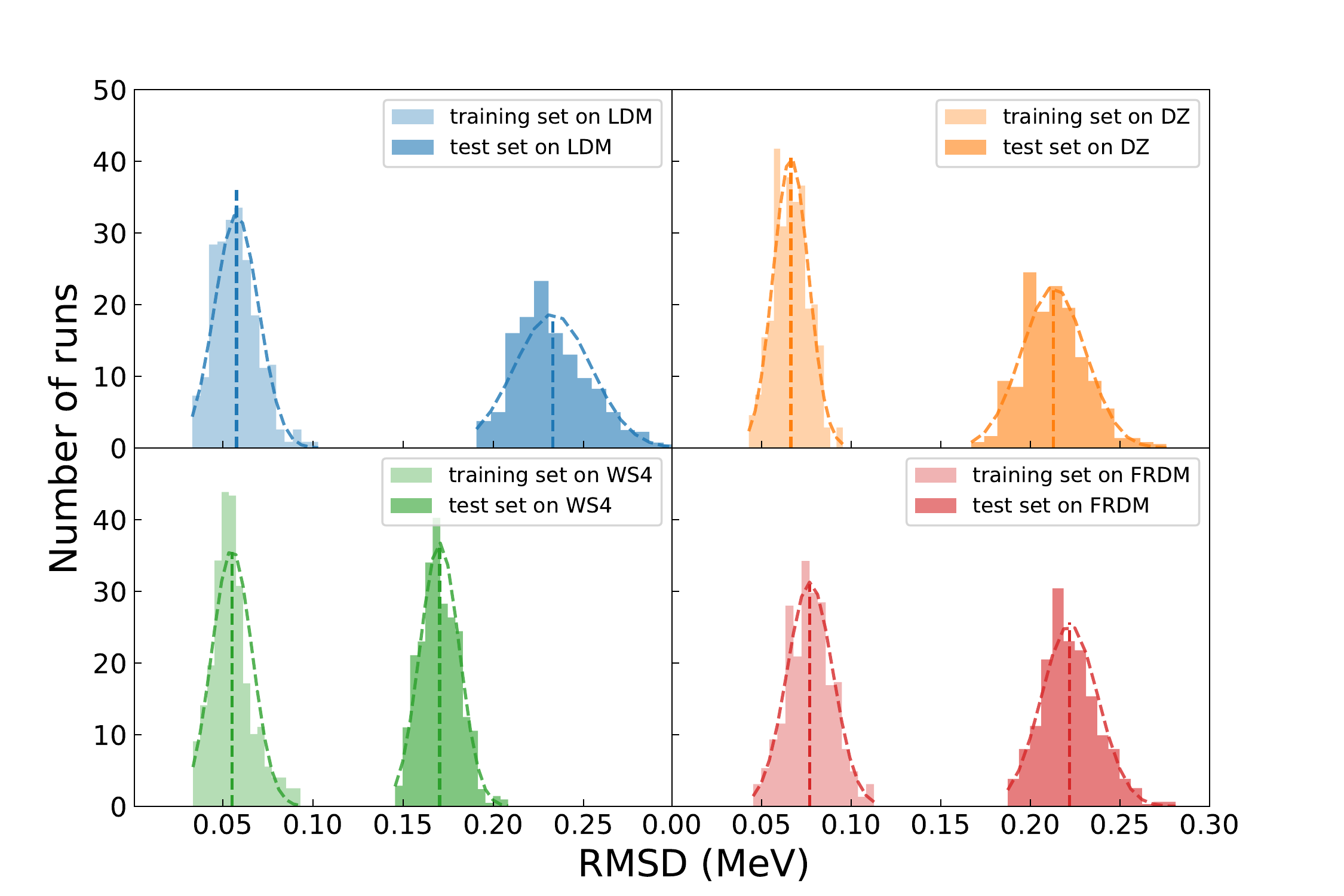}
    \caption{The density distribution of RMSD for training and testing data sets. Results from 500 runs for each mass model (LDM, DZ, WS4 and FRDM) are displayed. Dash lines denote Gaussian fit to the distribution. The corresponding mean value and the standard deviation are listed in Tab. \ref{different models}. In each run, the 2408 nuclei are randomly split into the training and testing data sets with the ratio of 4:1.}
    \label{fig:6}
\end{figure*}

\begin{table*}[!htb]
\caption{Comparison of the RMSD for the ML refined mass models. $\sigma_{pre}$ denotes the RMSD of the original mass models, $\sigma_{pre}$ is the result obtained with the LightGBM-refined mass models.  }
\label{different models}
\begin{tabular*}{16cm}{@{\extracolsep{\fill}} ccccccc}
\toprule
& & & LDM & DZ & WS4 & FRDM\\
\midrule

\multirow{6}{*}{Training set} & $\sigma_{pre}$ & & $2.462 \pm  0.023$ & $0.613 \pm  0.007$ & $0.302 \pm  0.003$ & $0.599 \pm  0.009$ \\
\cline{2-7}
& & LMNN by H. F. Zhang \cite{Zhang:2017zvb} & 0.235 & 0.325 & --- & 0.348 \\
& & BNN by Z. M. Niu \cite{Niu:2018csp} & --- & --- & 0.176 & 0.187 \\
& $\sigma_{post}$ & NN by R. Utama \cite{Utama:2015hva} & 0.466 & 0.274 & --- & 0.342 \\
& & NN by A. Pastore \cite{Pastore:2019lco} & --- & 0.324 & --- & ---  \\
& & Trees by M. Carnini \cite{Carnini:2020lvr} & 2.070 & 0.471 & --- & --- \\
& & LightGBM in this work & $0.058 \pm 0.011$ & $0.066 \pm 0.010$ & $0.055 \pm 0.011$ & $0.077 \pm 0.013$ \\

\midrule
\multirow{6}{*}{Testing set} & $\sigma_{pre}$ & & $2.467 \pm  0.092$ & $0.614 \pm  0.028$ & $0.303 \pm  0.011$ & $0.599 \pm  0.034$ \\
\cline{2-7}
& & LMNN by H. F. Zhang & 0.256 & 0.329 & --- & 0.368 \\
& & BNN by Z. M. Niu & --- & --- & 0.212 & 0.252 \\
& $\sigma_{post}$ & NN by R. Utama & 0.486 & 0.278 & --- & 0.352 \\
& & NN by A. Pastore & --- & 0.358 & --- & ---  \\
& & Trees by M. Carnini & 2.881 & 0.569 & --- & --- \\
& & LightGBM in this work & $0.234 \pm 0.022$ & $0.213 \pm 0.018$ & $0.170 \pm 0.011$ & $0.222 \pm 0.016$ \\
\bottomrule
\end{tabular*}
\end{table*}
In the previous section, the capability of LightGBM to refine LDM has been exhibited. In this section, besides LDM, three popular mass models, i.e., DZ, WS4 and FRDM, are tested as well. To do so, the $\delta(Z,A)$ between experimental binding energy and the one obtained from each mass model is fed to LightGBM, and we randomly split the 2408 nuclei into training and testing groups with the ratio of 4:1, and run 500 times for each mass model. The distribution of RMSD on the training and testing data sets are displayed in Fig.\ref{fig:6}. In Tab. \ref{different models}, the performance of serval ML refined mass models are compared. It can be seen that the typical value of RMSD on the training data set is only about 0.05-0.1 MeV, which is the smallest of all, to our best knowledge, the highest performance mass model. The typical value of RMSD on the testing data set is bout 0.2 MeV, which is also smaller than others. In general, significant improvements of about 90\%, 65\%, 40\%, and 60\% after the LightGBM refinement on the LDM, DZ, WS4, and FRDM are obtained, indicating the strong capability of LightGBM to improve theoretical nuclear mass models.

\begin{figure*}
    \centering
    \includegraphics[width=0.8\textwidth]{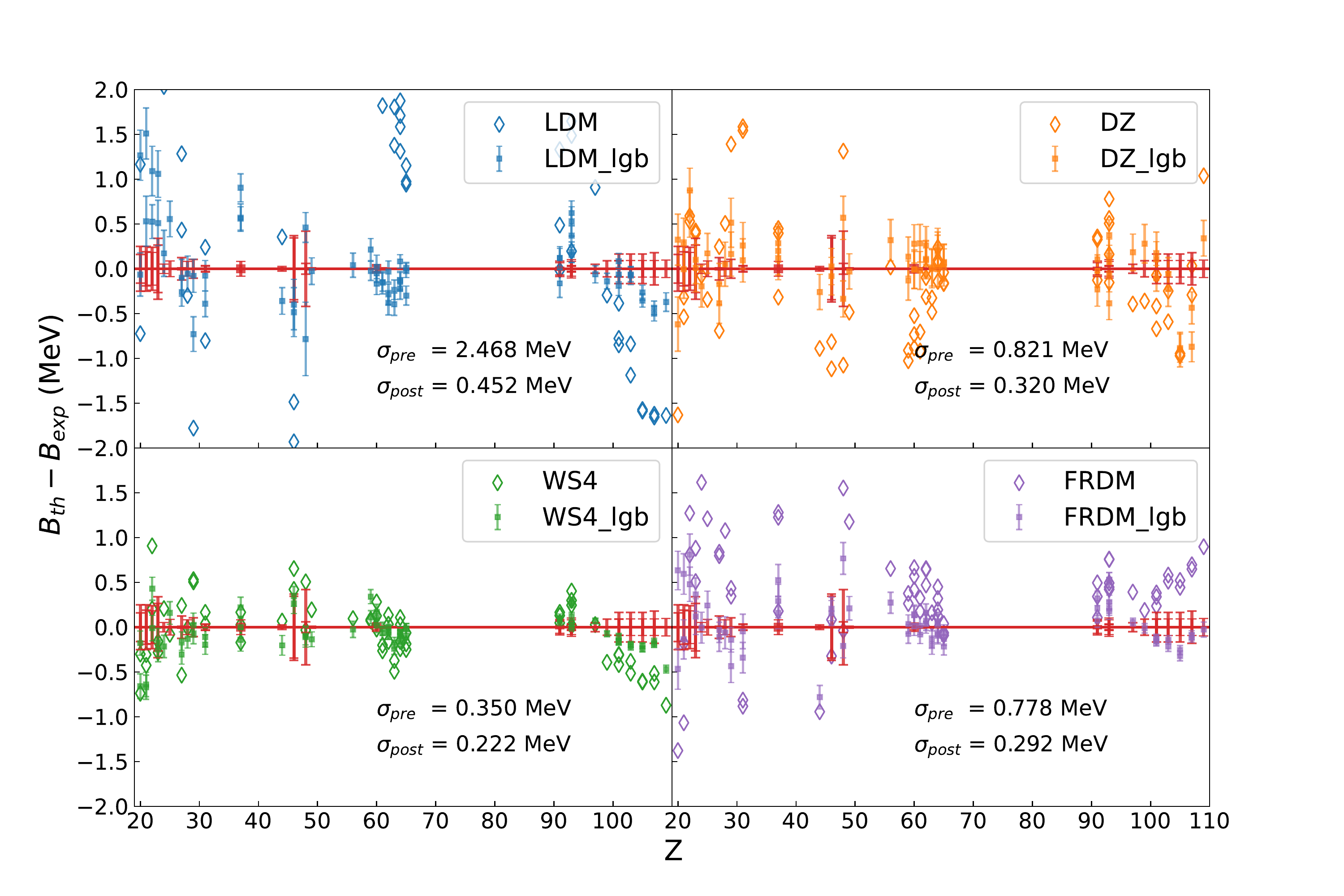}
    \caption{ The difference between the theoretical and the experimental binding energies (red horizontal line) obtained using LDM, DZ, WS4, and FRDM (open diamonds) and LightGBM-refined mass models (solid squares).  The results of 66 newly measured nuclei that appeared in the AME2020 mass evaluation are displayed. $\sigma_{pre}$ and $\sigma_{post}$ denote the RMSD of the original and the LightGBM-refined mass models on the newly measured nuclei, respectively. The error of the predictions obtained with the LightGBM-refined mass models is the standard deviation of the predicted binding energy. It is obtained by running LightGBM 500 times with randomly splitting AME2016 data into training and testing sets with the ratio of 4:1. }
    \label{fig:66}
\end{figure*}

Very recently, the AME2020 was published, thus it is interesting to see whether the LightGBM-refined mass models also work well for newly measured nuclei that appeared in the AME2020 mass evaluation.
The comparison of the binding energy obtained with LDM, DZ, WS4, and FRDM and LightGBM-refined mass models on the 66 newly measured nuclei that appeared in the AME2020 are illustrated in Fig.\ref{fig:66}. The RMSD of the original mass models on these newly measured nuclei are 2.468, 0.821, 0.350, and 0.778 MeV for the LDM, DZ, WS4, and FRDM, respectively. After the refinement of LightGBM, the RMSD of the above four mass models is significantly reduced to 0.452, 0.320, 0.222, and 0.292 MeV.

\subsection{Extrapolation of Neutron Separation Energy}

\begin{figure*}
    \centering
    \includegraphics[width=0.8\textwidth]{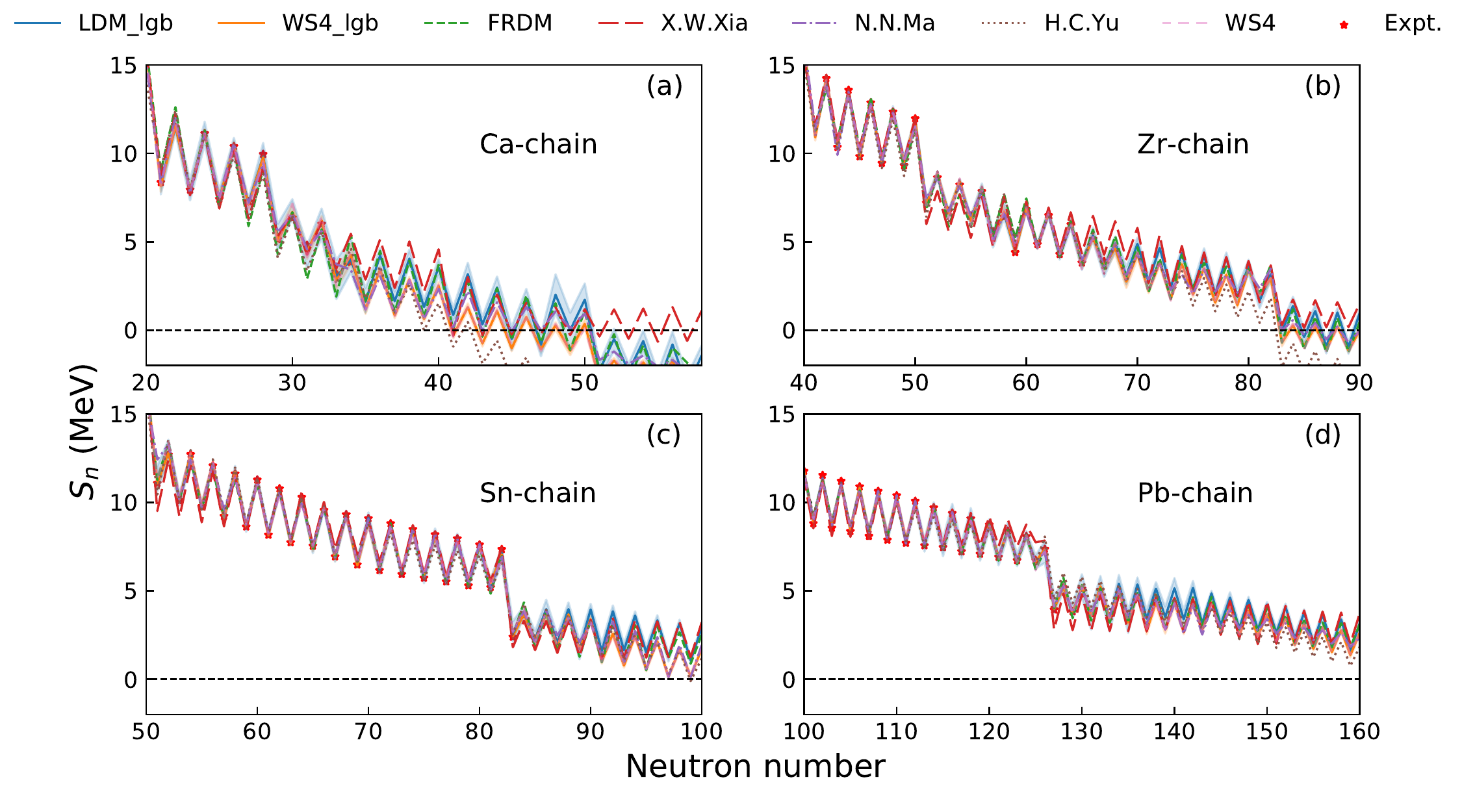}
    \caption{Single neutron separation energy of Ca, Zr, Sn, and Pb isotopic chains given by different models. The results obtained using LightGBM-refined LDM and WS4 are compared with FRDM, WS4, as well as recent theoretical calculations given by Xia et al. \cite{Xia:2017zka} , Ma et al. \cite{ma} , and Yu et al. \cite{yu}.}
    \label{fig:7}
\end{figure*}

\begin{figure*}
    \centering
    \includegraphics[width=0.8\textwidth]{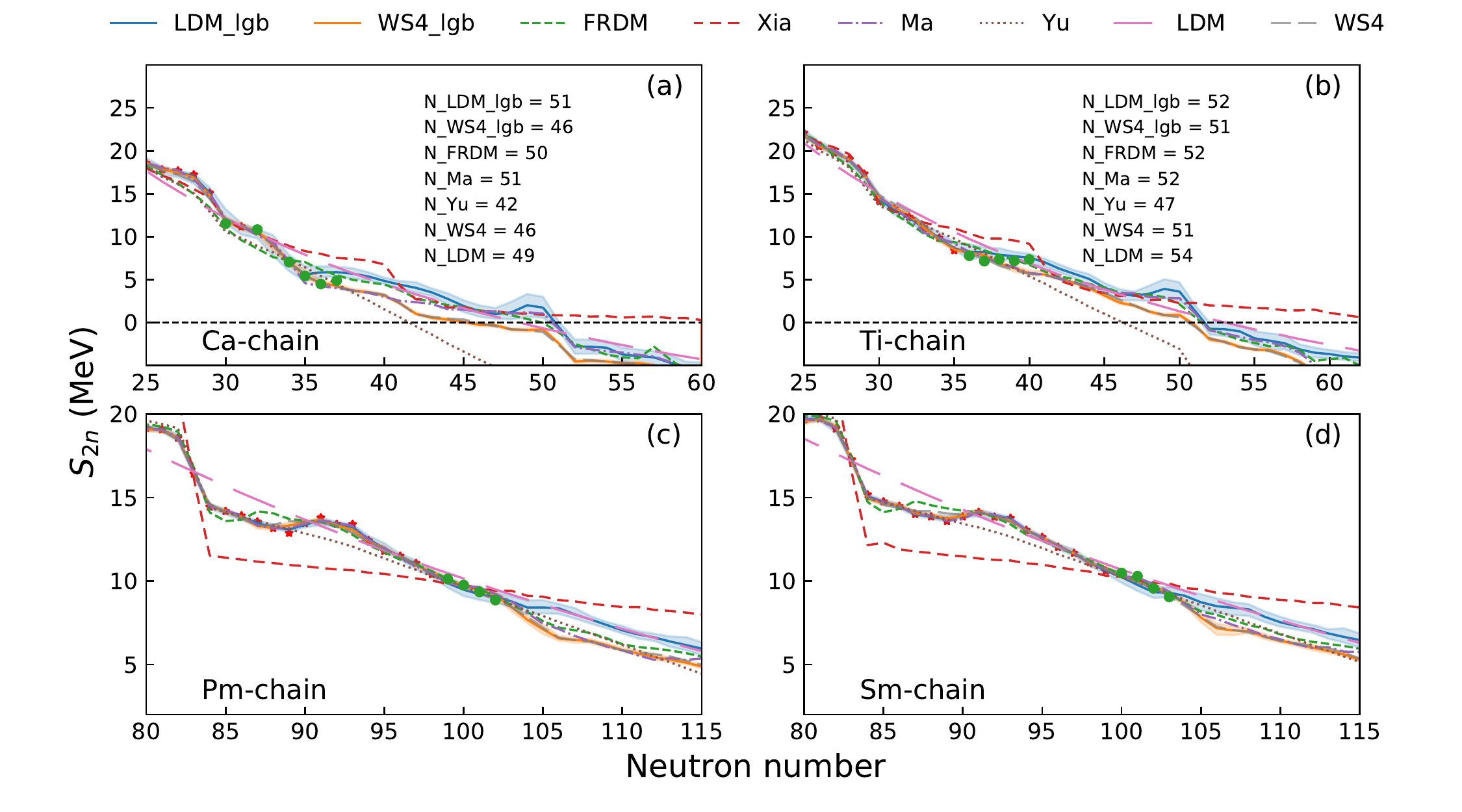}
    \caption{Two-neutron separation energy of the neutron-rich nuclei on Ca, Ti, Pm, Sm isotopic chains given by different models. The red stars and green dots represent the experimental data from AME2016 and the latest measurement from Refs.\cite{Michimasa:2020rbc,Vilen:2018ooh}. The neutron number of the predicted drip line isotopes (Ca and Ti) in each nuclear mass model are also listed in the figure. Note that $S_{2n}$ obtained with WS4 and LightGBM-refined WS4 are almost completely overlapped.}
    \label{fig:9}
\end{figure*}

Single and two-neutron separation energies are of particular interest, because they provide information relevant to shell and subshell structure, nuclear deformation, paring effects, as well as the boundary of the nuclear landscape. They can be calculated by the following formula:

\begin{equation}
\left\{
\begin{aligned}
&S_n(Z,A) = B(Z,A) - B(Z,A-1)\\
&S_{2n}(Z,A) = B(Z,A) - B(Z,A-2).\\
\end{aligned}
\right.
\end{equation}

Good performance of the LightGBM-refined mass models on the prediction of nuclear binding has been shown, it is interesting to see whether the single and two-neutron separation energies also can be reproduced well on the same foot. Fig.\ref{fig:7} compares the single neutron separation energy of Ca, Zr, Sn, and Pb isotopic chains given by different theoretical models and the experimental data from AME2016. All predictions are in good agreement with experimental data whenever there has data, while discrepancy appears as the increase of neutron number. The general trend of the $S_n$ as a function of neutron number obtained with LightGBM-refined LDM and WS4 are similar as that obtained with other nuclear mass models, e.g., the odd-even staggering also can be observed.

The latest experimental measurements of the two-neutron separation energy of the four elements (Ca, Ti, Pm, Sm) are compared with various theoretical model calculations in Fig.~\ref{fig:9}. It can be seen that the newly measured $S_{2n}$ can be well reproduced by LightGBM-refined LDM and WS4 models, particularly, $S_{2n}$ obtained with LightGBM-refined LDM lies much more closely to the experimental data than that obtained with LDM. For example, the sharp decrease of $S_{2n}$ around magic number cannot be reproduced by LDM, while this issue can be fixed after the refinement of LightGBM. Good performance of LightGBM-refined mass models on both $S_{n}$ and $S_{2n}$ indicating again the strong capacity of LightGMB on refining nuclear mass model.

\subsection{Interpretability of the model}

\begin{figure*}
    \centering
    \includegraphics[width=0.8\textwidth]{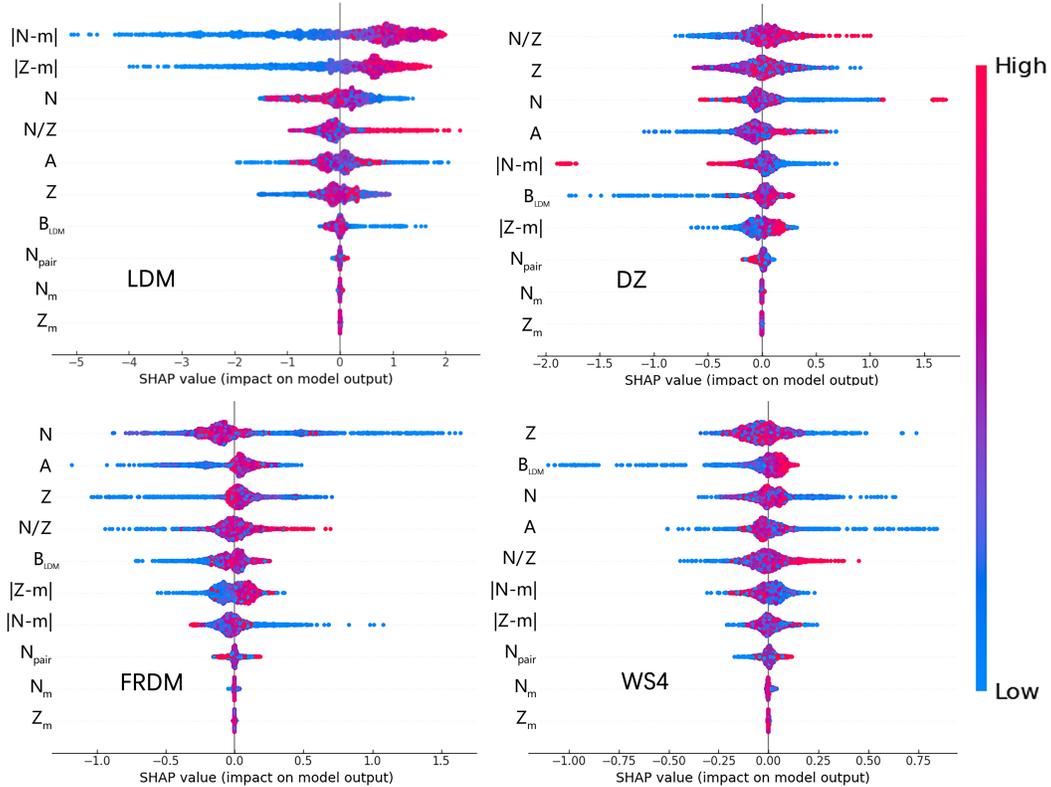}
    \caption{Importance ranking for the input features obtained with SHAP package. Each row represents a feature, and the x-axis is the SHAP value which shows how important a feature is for a particular prediction. Each point represents a nucleus, and the color represents the feature value (red is high, blue is low).}
    \label{fig:10}
\end{figure*}

\begin{figure}
    \centering
    \includegraphics[width=\linewidth]{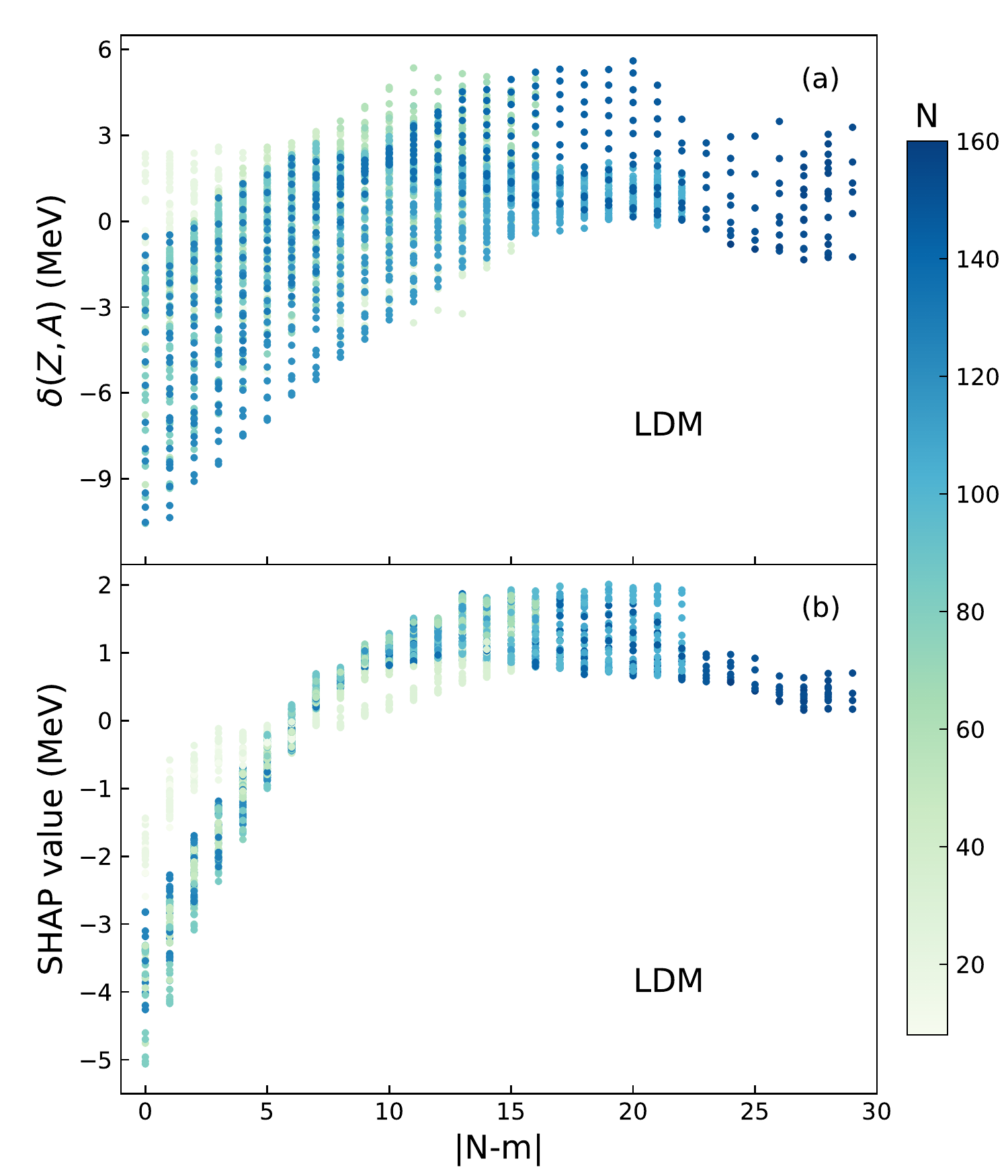}
    \caption{Upper: The residual $\delta(Z,A)$ obtained from LDM is plotted against $|N-m|$ colored by neutron number. Each point represents a nucleus, and the color represents the number of neutron in a nucleus. Lower: The same as the upper one but SHAP value is plotted instead of the residual $\delta(Z,A)$.}
    \label{fig:10}
\end{figure}

As a decision-tree based algorithm, one of advantage of LightGBM is its excellent degree of explainability. This is important because, as physicist, one expecting ML algorithm not only has a good performance on refining nuclear mass models, but also can provide some underlying physics that is absent from the original nuclear mass models. Understanding what happens when ML algorithms make predictions that could help us further improving our knowledge about the relationship between the input feature quantity and the predicted value. One of the possible way to understand how the LightGBM algorithm gives a particular prediction is to appreciate the most important features that drive the model. For this purpose, SHapley Additive exPlanation (SHAP) \cite{shap}, which is one of the most popular feature attribution methods, is applied to obtain the contributions of each feature value to the prediction. Fig.\ref{fig:10} illustrates the ranking of importance of the input 10 features. The top is the most important feature, while the bottom is the most irrelevant feature, to predict the residual $\delta(Z,A)$ between the experimental and theoretical binding energy. It can be seen that the importance ranks of input features are different for different mass models. Because shell effects are not included in LDM, the residual $\delta(Z,A)$ around magic numbers are usually larger (also can be seen in Fig.\ref{fig:5}). As a result, $|N-m|$ and $|Z-m|$ are more important to predict the $\delta(Z,A)$ between LDM calculation and experimental data. To demonstrate the meaning of SHAP value, residual $\delta(Z,A)$ obtained from LDM and SHAP value are shown in Fig.\ref{fig:10}. In the upper panel of Fig.\ref{fig:10}, around magic number, i.e., $|N-m|$ is close to 0, larger difference between LDM calculated and experimental binding energy is existed, especially for nuclei with larger neutron number. While very similar behavior for SHAP value can be seen in the lower panel. It implies that by adding a $|N-m|$-related term in the LDM, the accuracy of LDM on calculating nuclear binding energy can be improved to some extend.
For FRDM, the neutron number $N$ stands for the most relevant feature and the SHAP value for smaller $N$ is usually larger. Indeed, the residual $\delta(Z,A)$ for nuclei with smaller neutron number $N$ is larger has already been observed in FRDM paper, i.e., Fig. 6 of Ref. \cite{Moller:2015fba}. In addition, one sees that $N_{pair}$, $Z_m$, and $N_m$ are three of the most irrelevant features to predict the residual $\delta(Z,A)$, it means that the performance on predicting $\delta(Z,A)$ may not be influenced if they are removed from input features.

\section{Conclusion and outlook}
\label{Conclusion}

To summarize, several features are fed into the LightGBM algorithm to study the residual $\delta(Z,A)$ between the theoretical and experimental binding energies, it turns out that LightGBM algorithm can mimic the patterns of $\delta(Z,A)$ with high accuracy, so as to refine theoretical mass models. In this work, significant reductions on the RMSD of about 90\%, 65\%, 40\%, and 60\% after the LightGBM refinement on the LDM, DZ, WS4, and FRDM are obtained, indicating the strong capability of LightGBM to improve theoretical nuclear mass models. In addition, the RMSD for various mass models with respect to the 66 newly measured nuclei that appeared in AME2020 (compared with AME2016) is reduced on the same level as well. Furthermore, it is found that single and two-neutron separation energies obtained with the LightGBM-refined mass models are in good agreement with the newly appeared experimental data. By using the SHAP package, the most relevant input features to predict the residual $\delta(Z,A)$ for each mass model are found out, which may provide guidance for the further developments of nuclear mass models.

The good performance of machine learning method on refining the nuclear mass model gives us a new tool to further investigate other properties of nuclei that we are interested in, such as, superheavy nuclei, halo nuclei, and nuclei around drip-line. In addition, with the development of the interpretable machine learning methods, more physical hints can be obtained thereby improving our understanding of present nuclear models.

\begin{acknowledgments}
Fruitful discussions with Prof. Jie Meng, Prof. Hongfei Zhang, Prof. Yumin Zhao, Dr. Nana Ma are greatly appreciated. The authors acknowledge
support by the computing server C3S2 in Huzhou University. The work is supported in part by the National Science Foundation of China Nos. U2032145, 11875125, and 12047568,
and the National Key Research and Development Program of China under Grant No. 2020YFE0202002, and the ``Ten Thousand Talent Program" of Zhejiang province (No. 2018R52017). The mass table for the LightGBM-refined mass models is available in the Supplemental Material.
\end{acknowledgments}


\begin{thebibliography}{0}%
\makeatletter
\providecommand \@ifxundefined [1]{%
 \@ifx{#1\undefined}
}%
\providecommand \@ifnum [1]{%
 \ifnum #1\expandafter \@firstoftwo
 \else \expandafter \@secondoftwo
 \fi
}%
\providecommand \@ifx [1]{%
 \ifx #1\expandafter \@firstoftwo
 \else \expandafter \@secondoftwo
 \fi
}%
\providecommand \natexlab [1]{#1}%
\providecommand \enquote  [1]{``#1''}%
\providecommand \bibnamefont  [1]{#1}%
\providecommand \bibfnamefont [1]{#1}%
\providecommand \citenamefont [1]{#1}%
\providecommand \href@noop [0]{\@secondoftwo}%
\providecommand \href [0]{\begingroup \@sanitize@url \@href}%
\providecommand \@href[1]{\@@startlink{#1}\@@href}%
\providecommand \@@href[1]{\endgroup#1\@@endlink}%
\providecommand \@sanitize@url [0]{\catcode `\\12\catcode `\$12\catcode
  `\&12\catcode `\#12\catcode `\^12\catcode `\_12\catcode `\%12\relax}%
\providecommand \@@startlink[1]{}%
\providecommand \@@endlink[0]{}%
\providecommand \url  [0]{\begingroup\@sanitize@url \@url }%
\providecommand \@url [1]{\endgroup\@href {#1}{\urlprefix }}%
\providecommand \urlprefix  [0]{URL }%
\providecommand \Eprint [0]{\href }%
\providecommand \doibase [0]{http://dx.doi.org/}%
\providecommand \selectlanguage [0]{\@gobble}%
\providecommand \bibinfo  [0]{\@secondoftwo}%
\providecommand \bibfield  [0]{\@secondoftwo}%
\providecommand \translation [1]{[#1]}%
\providecommand \BibitemOpen [0]{}%
\providecommand \bibitemStop [0]{}%
\providecommand \bibitemNoStop [0]{.\EOS\space}%
\providecommand \EOS [0]{\spacefactor3000\relax}%
\providecommand \BibitemShut  [1]{\csname bibitem#1\endcsname}%
\let\auto@bib@innerbib\@empty
\end{thebibliography}%


\begin{thebibliography}{89}

\bibitem{1}
D.~Lunney, J.~M.~Pearson and C.~Thibault,
Rev. Mod. Phys. \textbf{75}, 1021-1082 (2003).
doi:10.1103/RevModPhys.75.1021

\bibitem{2}
K. Blaum, Phys. Rep. \textbf{425}, 1 (2006).

\bibitem{3}
F.~Wienholtz, D.~Beck, K.~Blaum, C.~Borgmann, M.~Breitenfeldt, R.~B.~Cakirli, S.~George, F.~Herfurth, J.~D.~Holt and M.~Kowalska, \textit{et al.}
Nature \textbf{498}, no.7454, 346-349 (2013)
doi:10.1038/nature12226

\bibitem{4}
K.~Blaum, J.~Dilling and W.~Nortershauser,
Phys. Scripta T \textbf{152}, 014017 (2013)
doi:10.1088/0031-8949/2013/T152/014017
[arXiv:1210.4045 [physics.atom-ph]].

\bibitem{Niu:2018olp}
Z.~Niu, H.~Liang, B.~Sun, Y.~Niu, J.~Guo and J.~Meng,
Sci. Bull. \textbf{63}, 759-764 (2018)
doi:10.1016/j.scib.2018.05.009
[arXiv:1807.05535 [nucl-th]].
\bibitem{44}
Wang M, Zhang Y H, Zhou X H. Nuclear mass measurements (in Chinese). Sci Sin-Phys Mech Astron, 2020, 50: 052006, doi: 10.1360/SSPMA-2019-
0308

\bibitem{5}
C.~Ma, M.~Bao, Z.~M.~Niu, Y.~M.~Zhao and A.~Arima,
Phys. Rev. C \textbf{101}, no.4, 045204 (2020)
doi:10.1103/PhysRevC.101.045204

\bibitem{6}
M.~Wang, G,~Audi, F.~G.~Kondev, W.~J.~Huang, S.~Naimi, and X.~Xu,
Chinese Physics C(2017).

\bibitem{7}
P.~M\"oller, W.~D.~Myers, H.~Sagawa and S.~Yoshida,
Phys. Rev. Lett. \textbf{108}, no.5, 052501 (2012)
doi:10.1103/PhysRevLett.108.052501


\bibitem{8}
P.~M\"oller, J.~R.~Nix, W.~D.~Myers and W.~J.~Swiatecki,
Atom. Data Nucl. Data Tabl. \textbf{59}, 185-381 (1995)
doi:10.1006/adnd.1995.1002
[arXiv:nucl-th/9308022 [nucl-th]].

\bibitem{9}
N.~Wang, M.~Liu, X.~Wu and J.~Meng,
Phys. Lett. B \textbf{734}, 215-219 (2014)
doi:10.1016/j.physletb.2014.05.049
[arXiv:1405.2616 [nucl-th]].

\bibitem{10}
S.~Goriely, N.~Chamel and J.~M.~Pearson,
Phys. Rev. C \textbf{93}, no.3, 034337 (2016)
doi:10.1103/PhysRevC.93.034337

\bibitem{11}
S.~Goriely, N.~Chamel and J.~M.~Pearson,
Phys. Rev. Lett. \textbf{102}, 152503 (2009)
doi:10.1103/PhysRevLett.102.152503
[arXiv:0906.2607 [nucl-th]].

\bibitem{12}
Y.~Aboussir, J.~M.~Pearson, A.~K.~Dutta and F.~Tondeur,
Atom. Data Nucl. Data Tabl. \textbf{61}, 127-176 (1995)
doi:10.1016/S0092-640X(95)90014-4


\bibitem{Geng:2005yu}
L.~S.~Geng, H.~Toki and J.~Meng,
Prog. Theor. Phys. \textbf{113}, 785-800 (2005)
doi:10.1143/PTP.113.785
[arXiv:nucl-th/0503086 [nucl-th]].


\bibitem{Xia:2017zka}
X.~W.~Xia, Y.~Lim, P.~W.~Zhao, H.~Z.~Liang, X.~Y.~Qu, Y.~Chen, H.~Liu, L.~F.~Zhang, S.~Q.~Zhang and Y.~Kim, \textit{et al.}
Atom. Data Nucl. Data Tabl. \textbf{121-122}, 1-215 (2018)
doi:10.1016/j.adt.2017.09.001
[arXiv:1704.08906 [nucl-th]].





\bibitem{ROMP}
  C.~Giuseppe, C.~Ignacio, C.~Kyle {\it et al.},
  Rev.\ Mod.\ Phys.\ {\bf 91}, 045002 (2019).

\bibitem{Nature1}
  Radovic.~A, Williams, Rousseau.~D {\it et al.},
  Nature {\bf 560}, 41 (2018).





\bibitem{Utama:2015hva}
R.~Utama, J.~Piekarewicz and H.~B.~Prosper,
Phys. Rev. C \textbf{93}, no.1, 014311 (2016)
doi:10.1103/PhysRevC.93.014311
[arXiv:1508.06263 [nucl-th]].
\bibitem{Utama:2017wqe}
R.~Utama and J.~Piekarewicz,
Phys. Rev. C \textbf{96}, no.4, 044308 (2017)
doi:10.1103/PhysRevC.96.044308
[arXiv:1704.06632 [nucl-th]].

\bibitem{Niu:2018csp}
Z.~M.~Niu and H.~Z.~Liang,
Phys. Lett. B \textbf{778}, 48-53 (2018)
doi:10.1016/j.physletb.2018.01.002
[arXiv:1801.04411 [nucl-th]].



\bibitem{Utama:2016tcl}
R.~Utama, W.~C.~Chen and J.~Piekarewicz,
J. Phys. G \textbf{43}, no.11, 114002 (2016).

\bibitem{PJC}
  Z.~A.~Wang, J.~C.~Pei, Y.~Liu,
  Phys.\ Rev.\ Lett {\bf 123}, 122501 (2019).


\bibitem{Ma:2020bic}
C.~W.~Ma, D.~Peng, H.~L.~Wei, Y.~T.~Wang and J.~Pu,
Chin. Phys. C \textbf{44}, no.12, 124107 (2020).

\bibitem{Ma:2020mbd}
C.~W.~Ma, D.~Peng, H.~L.~Wei, Z.~M.~Niu, Y.~T.~Wang and R.~Wada,
Chin. Phys. C \textbf{44}, no.1, 014104 (2020).



\bibitem{Niu:2018trk}
Z.~M.~Niu, H.~Z.~Liang, B.~H.~Sun, W.~H.~Long and Y.~F.~Niu,
Phys. Rev. C \textbf{99}, no.6, 064307 (2019)
doi:10.1103/PhysRevC.99.064307
[arXiv:1810.03156 [nucl-th]].




\bibitem{PLG}
  L.~G.~Pang, K.~Zhou, N.~Su {\it et al.},
  Nat.\ Commun {\bf 9} 210 (2018).
\bibitem{Du:2019civ}
Y.~L.~Du, K.~Zhou, J.~Steinheimer, L.~G.~Pang, A.~Motornenko, H.~S.~Zong, X.~N.~Wang and H.~St\"ocker,
Eur. Phys. J. C \textbf{80}, no.6, 516 (2020)
doi:10.1140/epjc/s10052-020-8030-7
[arXiv:1910.11530 [hep-ph]].

\bibitem{Steinheimer:2019iso}
J.~Steinheimer, L.~Pang, K.~Zhou, V.~Koch, J.~Randrup and H.~Stoecker,
JHEP \textbf{12}, 122 (2019).
\bibitem{Song:2021rmm}
Y.~D.~Song, R.~Wang, Y.~G.~Ma, X.~G.~Deng and H.~L.~Liu,
Phys. Lett. B \textbf{814}, 136084 (2021)
doi:10.1016/j.physletb.2021.136084
[arXiv:2101.10613 [nucl-th]].
\bibitem{Wang:2020tgb}
R.~Wang, Y.~G.~Ma, R.~Wada, L.~W.~Chen, W.~B.~He, H.~L.~Liu and K.~J.~Sun,
Phys. Rev. Res. \textbf{2}, no.4, 043202 (2020)

\bibitem{Li:2020qqn}
F.~Li, Y.~Wang, H.~L\"u, P.~Li, Q.~Li and F.~Liu,
J. Phys. G \textbf{47}, no.11, 115104 (2020).

\bibitem{Zhang:2017zvb}
H.~F.~Zhang, L.~H.~Wang, J.~P.~Yin, P.~H.~Chen and H.~F.~Zhang,
J. Phys. G \textbf{44}, no.4, 045110 (2017)
doi:10.1088/1361-6471/aa5d78


\bibitem{shelley}
M. Shelley and A. Pastore, arXiv:2102.07497 (2021).

\bibitem{Neufcourt:2018syo}
L.~Neufcourt, Y.~Cao, W.~Nazarewicz and F.~Viens,
Phys. Rev. C \textbf{98}, no.3, 034318 (2018)
doi:10.1103/PhysRevC.98.034318
[arXiv:1806.00552 [nucl-th]].

\bibitem{Carnini:2020lvr}
M.~Carnini and A.~Pastore,
J. Phys. G \textbf{47}, no.8, 082001 (2020)
doi:10.1088/1361-6471/ab92e3
[arXiv:2002.10290 [nucl-th]].

\bibitem{MLP}
Esra Y\"uksel, Derya Soydaner, and H\"useyin Bahtiyar, arXiv:2101.12117v1 (2021).

\bibitem{Gernoth:1993dqa}
K.~A.~Gernoth, J.~W.~Clark, J.~S.~Prater and H.~Bohr,
Phys. Lett. B \textbf{300}, 1-7 (1993)
doi:10.1016/0370-2693(93)90738-4

\bibitem{Athanassopoulos:2003qe}
S.~Athanassopoulos, E.~Mavrommatis, K.~A.~Gernoth and J.~W.~Clark,
Nucl. Phys. A \textbf{743}, 222-235 (2004)
doi:10.1016/j.nuclphysa.2004.08.006
[arXiv:nucl-th/0307117 [nucl-th]].

\bibitem{Clark:2006ua}
J.~W.~Clark and H.~Li,
Int. J. Mod. Phys. B \textbf{20}, no.30n31, 5015-5029 (2006)
doi:10.1142/S0217979206036053
[arXiv:nucl-th/0603037 [nucl-th]].



\bibitem{Duflo:1995ep}
J.~Duflo and A.~P.~Zuker,
Phys. Rev. C \textbf{52}, R23 (1995)
doi:10.1103/PhysRevC.52.R23
[arXiv:nucl-th/9505011 [nucl-th]].

\bibitem{Pastore:2019lco}
A.~Pastore, D.~Neill, H.~Powell, K.~Medler and C.~Barton,
Phys. Rev. C \textbf{101}, no.3, 035804 (2020)
doi:10.1103/PhysRevC.101.035804
[arXiv:1912.11365 [nucl-th]].

\bibitem{ma}
N.~N.~Ma, H.~F.~Zhang, X.~J.~Bao and H.~F.~Zhang,
Chin. Phys. C \textbf{43}, no.4, 044105 (2019)
doi:10.1088/1674-1137/43/4/044105


\bibitem{yu}
H.~C.~Yu, M.~Q.~Lin, M.~Bao, Y.~M.~Zhao and A.~Arima,
Phys. Rev. C \textbf{100}, no.1, 014314 (2019)
doi:10.1103/PhysRevC.100.014314


\bibitem{Michimasa:2020rbc}
S.~Michimasa, M.~Kobayashi, Y.~Kiyokawa, S.~Ota, R.~Yokoyama, D.~Nishimura, D.~S.~Ahn, H.~Baba, G.~P.~A.~Berg and M.~Dozono, \textit{et al.}
Phys. Rev. Lett. \textbf{125}, no.12, 122501 (2020)
doi:10.1103/PhysRevLett.125.122501

\bibitem{Vilen:2018ooh}
M.~Vilen, J.~M.~Kelly, A.~Kankainen, M.~Brodeur, A.~Aprahamian, L.~Canete, T.~Eronen, A.~Jokinen, T.~Kuta and I.~D.~Moore, \textit{et al.}
Phys. Rev. Lett. \textbf{120}, no.26, 262701 (2018)
[erratum: Phys. Rev. Lett. \textbf{124}, no.12, 129901 (2020)]
doi:10.1103/PhysRevLett.120.262701
[arXiv:1801.08940 [nucl-ex]].

\bibitem{shap}
  S.~Lundberg, S.~I.~Lee.
  arXiv:1705.07874 (2017) [cs.AI].

\bibitem{Moller:2015fba}
P.~M\"oller, A.~J.~Sierk, T.~Ichikawa and H.~Sagawa,
Atom. Data Nucl. Data Tabl. \textbf{109-110}, 1-204 (2016)
doi:10.1016/j.adt.2015.10.002
[arXiv:1508.06294 [nucl-th]].

\bibitem{LightGBM}
Guolin Ke, Qi Meng, Thomas Finley, Taifeng Wang, Wei Chen, Weidong Ma, Qiwei Ye, Tie-Yan Liu.
``LightGBM: A Highly Efficient Gradient Boosting Decision Tree.''
Advances in Neural Information Processing Systems 30 (NIPS 2017), pp. 3149-3157.

\end{thebibliography}
\end{document}